\documentclass[preprint,superscriptaddress,keywords,preprintnumbers,amsmath,amssymb,aps,prb]{revtex4}
\usepackage{graphicx}
\usepackage{dcolumn}
\usepackage{bm}

\begin{document}
\title{Application of the Huang-Hilbert transform and natural time to the analysis of Seismic Electric Signal activities}
\author{K. A. Papadopoulou}
\affiliation{Department of Solid State Physics and Solid Earth
Physics Institute, Faculty of Physics, School of Science, National
and Kapodistrian University of Athens, Panepistimiopolis, Zografos
157 84, Athens, Greece}
\author{E. S. Skordas}
\email{eskordas@phys.uoa.gr} \affiliation{Department of Solid
State Physics and Solid Earth Physics Institute, Faculty of
Physics, School of Science, National and Kapodistrian University
of Athens, Panepistimiopolis, Zografos 157 84, Athens, Greece}

\begin{abstract}
The Huang-Hilbert transform is applied to Seismic Electric Signal
(SES) activities in order to decompose them into a number of
Intrinsic Mode Functions (IMFs) and study which of these functions
better represent the SES. The results are compared to those
obtained from the analysis in a new time domain termed natural
time after having subtracted the magnetotelluric background from
the original signal. It is shown that the instantaneous amplitudes
of the IMFs can be used for the distinction of SES from artificial
noises when combined with the natural time analysis.

{\bf Keywords:} seismic electric signal, natural time, Huang-Hilbert transform, instantaneous amplitude
\end{abstract}
\maketitle

{\bf Seismic Electric Signal (SES) activities are low frequency
electric signals precursory to earthquakes that have been
found to exhibit critical dynamics. Their distinction from
noise which is usually characterized by different dynam-
ics is an important task. Here, the Huang decomposition
method and the Hilbert transform, used for the analysis
of non-stationary time series, have been applied for the
first time to time-series of SES activities of short duration
in order to study their features. By decomposing the SES
activity using the Huang method into some functions,
called Intrinsic Mode Functions (IMFs), it is found that
the first of the IMFs, hereafter IMF1, does not contain
enough information to secure a classification of the signal
(i.e., SES or noise), but it might contain some information
about the distance of the source emitting the electric sig-
nal, from the recording station. Furthermore, the Hilbert
transform is applied to the IMFs in order to determine
their  instantaneous  amplitudes  $A(t)$. Earlier  studies
revealed that the classification of the signal as an SES can
be achieved upon employing natural time analysis and
the instantaneous power $P$ of the time-series as described
in Ref. \onlinecite{NAT09V}, provided that the usual sinusoidal background
has been first determined and then eliminated. Here however, it is shown that the squared instantaneous amplitude  $A(t)^2$ of the sum of all the IMFs (i.e., the $A^2$ of the
original time-series, including the background) can be used directly for the analysis in natural time in order to classify the signal as an SES.}

\section{Introduction}

Seismic Electric Signals (SES)\cite{VAR84A,VAR84B,NEWBOOK}
are low frequency $(\leq1 Hz)$ changes of the electric field of
the earth, that have been found to precede earthquakes with lead
times ranging from several hours to a few
months\cite{UYE96,VAR91}. They are emitted when the gradually
increasing stress before an earthquake reaches a critical value
\cite{VAR93} in which the electric dipoles formed due to point defects\cite{varotsos1974new,LAZ85,KOS75}
in the future focal area that affect the usual dielectric properties\cite{VARALEX82B} exhibit cooperative orientation, thus leading to an emission of a transient electric signal.
Upon analysing these signals in a new time domain
termed natural time\cite{NAT01, NAT02}, they can be
distinguished\cite{NAT03A, NAT03B} from other electric signals
of different origin, e.g., signals emitted from man-made
electrical sources. The latter are hereafter called ``artificial''
signals.

In general, the analysis in natural time may uncover properties
hidden in complex time-series\cite{ABE05}. In particular, for
the case of SES, the natural time analysis of the small
earthquakes subsequent to a series of SES, termed SES
activity\cite{VAR91}, may identify the occurrence time of the
impending mainshock\cite{SAR08}.

For a time-series consisting of \emph{N} events, the natural time
can be defined\cite{SPRINGER} as  $\chi_k = k/N$ and it serves as
an index for the occurrence of the \emph{k}-th event. We
study\cite{NAT02, NAT06B, NAT05C, NAT03A, NAT04, PNAS}
the evolution of the pair ($\chi_k, Q_k$) where $Q_k$ is a
quantity proportional to the energy released in the \emph{k}-th
event. For a signal of dichotomous nature, $Q_k$ is proportional
to the duration of the \emph{k}-th event; otherwise, the quantity
$Q_k$ is determined\cite{NAT09V} by means of the instantaneous
power $P$.

The normalized power spectrum is introduced as\cite{NAT01}
$\Pi(\omega )=|\Phi (\omega )|^2$ where
\begin{equation}
\label{eq:1}
\Phi(\omega)=\sum_{k=1}^{N} p_k \exp (i\omega\frac{k}{N})
\end{equation}
with $p_k=Q_{k}/\sum_{n=1}^{N}Q_{n}$. As $\omega \rightarrow 0$
Eq.~\eqref{eq:1} gives some useful statistical properties. For
example, it has been shown\cite{NAT01} that
\begin{equation}
\label{eq:2}
\Pi(\omega)=1-0.07\omega^2+\ldots
\end{equation}
which for an SES activity leads to\cite{NAT01} the variance of
$\chi$ hereafter labelled $\kappa_1$:
\begin{equation}
\label{eq:3}
\kappa_1=<\chi^2>-<\chi>^2=0.07
\end{equation}
Other parameters we take under consideration are the entropy
\emph{S} in natural time given by\cite{NAT03B}
\begin{equation}
\label{eq:4}
S=<\chi\ln{\chi}>-<\chi>\ln{<\chi>}
\end{equation}
where $<f(\chi)>\equiv\sum_{k=1}^{N}p_k f(\chi_{k})$, and the
entropy under time reversal $S_-\equiv\hat{T}S$, where the effect
of the time-reversal operator $\hat{T}$ on $Q_k$ is given
by\cite{SPRINGER} $\hat{T}Q_k=Q_{N-k+1}$ which positions the first
pulse (k=1) as last in the new time reversed time-series etc. It
has been shown\cite{NAT03B} that an SES activity obeys the
inequalities:
\begin{equation}
\label{eq:5}
S,S_-<S_u
\end{equation}
where $S_u$ the entropy  of the uniform distribution $S_u=0.0966$.
The relations \eqref{eq:3} and \eqref{eq:5}, which involve the
parameters $\kappa_1$, $S$ and $S_-$, are used in order to
classify as SES an electric signal of short duration after
eliminating its sinusoidal background, termed
magnetotelluric\cite{VAR03, SAR02, NEWBOOK} background.
This background comprises electrical variations induced by
frequent small variations of the Earth's magnetic field originated
from extraterrestrial sources. An example of an electric signal
analysis following the procedure described above will be given
below in Section III.

The Huang-Hilbert transform\cite{huang1998, huang2003, yang2004, bowman2013,
xie2006} shortly reviewed in Section \ref{hh} is a novel method used
in the analysis of non-stationary time-series by computing
instantaneous amplitudes and frequencies. One of its components is
the Empirical Mode Decomposition (EMD) which decomposes a signal
into some functions called Intrinsic Mode Functions (IMF) that, in
combination with the Hilbert transform, can give information about
the spectrum of the signal. It is the scope of  the present paper
to study the application of the Huang-Hilbert
transform\cite{huang1998, huang2003, yang2004, bowman2013,
xie2006} to SES
activities of short duration by considering also natural time
analysis. In particular, the latter analysis is used for the
classification of an SES activity in Section \ref{nt}, while the
IMFs of the SES activity are extracted in Section \ref{hhapp}. The
interconnection of these results is investigated in Section
\ref{instamp}. Specifically, in order to determine which of the
IMFs represent the SES, we show in Section V, that the squared
instantaneous amplitudes as derived after the application of the
Hilbert transform can also be used instead of the instantaneous
power for the classification of the electric signal as SES. The
reason we choose to study the instantaneous amplitudes is because,
as shown in Section V (in particular see
Fig.~\ref{fig:sumsA}c that will be discussed later), the
instantaneous amplitude of the sum of all the IMFs extracted from
the signal can approximate the original time-series.

\begin{figure}
\includegraphics[scale=0.5]{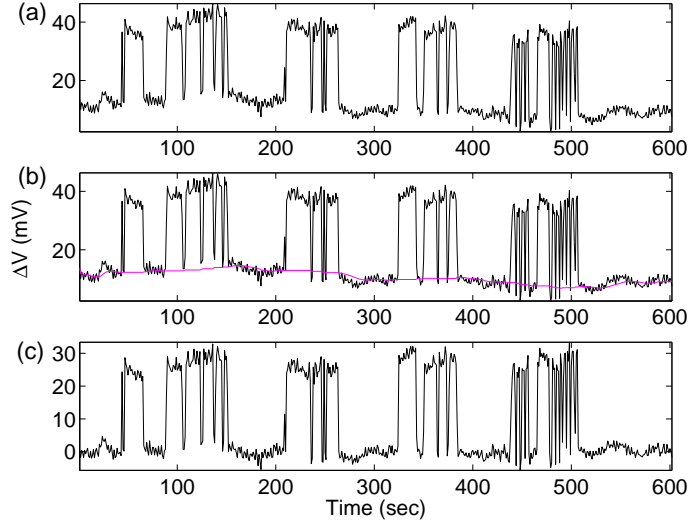}
\caption{\label{fig:shortses1} (color online) a)The short duration
electric signal used in this analysis, b) The signal with its
background (magenta solid line) after the application of the
exponential fit, and c)The signal after the elimination of the
background.}
\end{figure}

\begin{figure}
\includegraphics[scale=0.5]{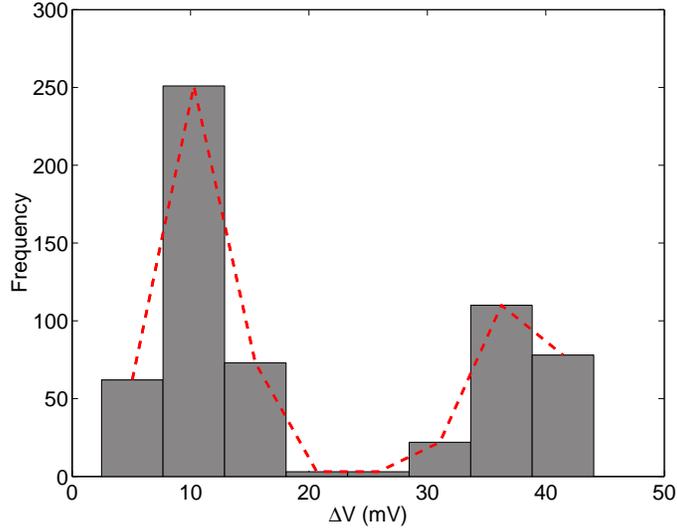}
\caption{\label{fig:histpdf} (color online) The histogram of the
signal of Fig.~\ref{fig:shortses1}a and its pdf.}
\end{figure}

\section{The Huang-Hilbert Transform}\label{hh}

The Hilbert transform of a function $x(t)$ is\cite{huang1998}:
\begin{equation}
\label{eq:8}
H[x(t)]\equiv\hat{y}(t)=\frac{1}{\pi}PV\int_{-\infty}^{+\infty}\frac{x(\tau)}{t-\tau}d\tau
\end{equation}
where $PV$ indicates the Cauchy principal value. The original time-series and its Hilbert transform are the real and imaginary part respectively of an analytic function $z(t)$
\begin{equation}
\label{eq:9}
z(t)=x(t)+i\hat{y}(t)=A(t)e^{i\theta(t)}
\end{equation}
where
\begin{equation}
\label{eq:10}
A(t)=(x^2+{\hat{y}}^2)^\frac{1}{2}
\end{equation}
stand for the instantaneous amplitudes.

An IMF should satisfy two conditions\cite{huang1998}: First, the
number of extrema and the number of zero crossings must be equal
or differ at most by one. Second, the mean value of the envelopes
defined by the local maxima and minima is zero. If we apply the
Hilbert transform to the IMFs  we derive the instantaneous
amplitudes $A(t)$ which will be used later for the analysis of the
signal in natural time.

In order to begin the EMD (the process which extracts the IMFs) the signal has to have at least two extrema\cite{huang1998}. If there are not any extrema then the signal can be differentiated to reveal them. Finally, the characteristic time scale is defined by the time lapse between the extrema and not the zero crossings. The decomposition, i.e. the method to extract the IMFs, is designated as the sifting process and is described below\cite{huang2003, yang2004, bowman2013,
xie2006}:\\
a) The maxima are determined and connected with a cubic spline, thus creating an upper envelope.\\
b) The minima are determined and connected with a cubic spline so that a lower envelope is created. These two envelopes should contain all the data.\\
c) The mean value of the envelopes $m=\big((upper-envelope)+(lower-envelope)\big)/2$ is calculated and subtracted from the signal $X(t)$ deriving a component $h$, $X(t)-m=h(t)$.\\
d) The sifting process starts anew treating the component $h$ as the data and a new component $h'$ is derived etc.\\
This process will stop when one of the $h$ components satisfies the definition of an IMF (zero mean value, the number of extrema and zero crossings is equal or differ at most by one). The first IMF will be the last $h$ component extracted. We set $h=c$.\\
e) The IMF c is subtracted from the original signal and the process begins anew (steps a-d) treating the residual $res= X(t)-c$ as the data until the second IMF is extracred etc.\\
The original signal can be recomposed using the following relation
\begin{equation}
\label{eq:11}
X(t)=\sum_{i=1}^n c_i + trend
\end{equation}
where \emph{trend} stands for the last IMF. When the final residual is a monotonic function the process stops.

\section{Natural time analysis of an SES activity of short duration}\label{nt}

Figure \ref{fig:shortses1}a depicts the electric signal that is
used as a typical example in the present analysis. It was recorded
at a measuring station located close to Volos city, Greece, on
November 21, 2012 and preceded the $M_b$=4.6 earthquake with an
epicentre at 38.26$^o$N,  22.22$^o$E that occurred on April 28,
2013.

\begin{figure}
\includegraphics[scale=0.5]{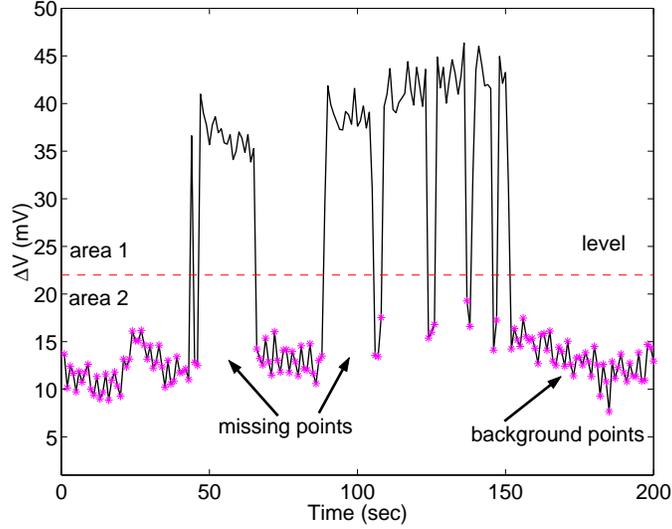}
\caption{\label{fig:level} (color online) The level which
separates the high-level from the low-level states is represented
with the horizontal dotted line. The points below the level
(asterisks in area 2) are considered to be the magnetotelluric
background. The ``missing points'', will be re-calculated using an
exponential fit(see the text).}
\end{figure}

\begin{figure}
\includegraphics[scale=0.5]{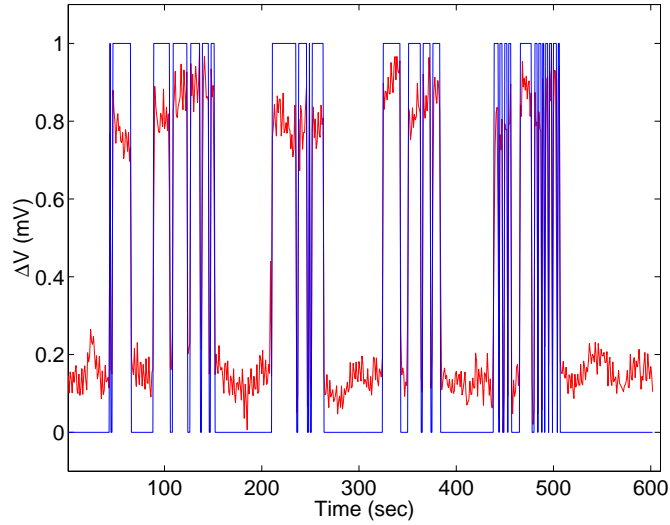}
\caption{\label{fig:dichotomous} (color online) The dichotomous
signal (blue) and the original signal (red) after the rescaling.}
\end{figure}

\begin{table}
\caption{\label{tab:sumsimf}Analysis in natural time of IMF sums.}
\begin{tabular}{cccc}
\hline
Sums of IMF: & $\kappa_1$ & $S$ & $S_-$ \\
\hline
1 & $0.077\pm0.008$ & $0.087\pm0.012$ & $0.086\pm0.018$\\
1-3 & $0.074\pm0.003$ & $0.073\pm0.005$ & $0.097\pm0.003$\\
1-5 & $0.081\pm0.004$ & $0.104\pm0.006$ & $0.077\pm0.005$\\
1-9 & $0.060\pm0.012$ & $0.075\pm0.014$ & $0.052\pm0.017$\\
2-4 & $0.080\pm0.003$ & $0.080\pm0.005$ & $0.098\pm0.002$\\
2-9 & $0.062\pm0.014$ & $0.078\pm0.012$ & $0.056\pm0.016$\\
3-9 & $0.056\pm0.006$ & $0.069\pm0.004$ & $0.050\pm0.010$\\
\hline
\end{tabular}
\end{table}

In order to eliminate the magnetotelluric background we work as follows:\\
First, we plot the histogram of the signal along with its
probability density function (pdf) see Fig.~\ref{fig:histpdf}. We
need to determine a level which will separate the high-level
states (i.e., those having the largest deflections of the electric
field amplitude) from the low-level states designated as the
background\cite{NAT03A}. For this purpose, the minimum value of
the pdf, i.e., 22mV in this case, (Fig.~\ref{fig:histpdf}) is
selected as background level. Thus, two time series of
subsequently high-level and low-level states are constructed. The
values greater than 22mV (the high-level states) will represent
the signal while the values below this level (i.e. the low-level
states) correspond to the magnetotelluric background.

The time-series of the background appears to have some discontinuities (``missing points'' in Fig.~\ref{fig:level}). We re-calculate these ``missing points'' following the process below:\\
The mean duration of an SES pulse lies in the range\cite{SPRINGER}
11-14sec. In every ``missing point'' of Fig.~\ref{fig:level} a new
value is assigned. We assume that this value is a past average
with weight $e^{(-\frac{\Delta t}{20})}$. Thus, we employ the
following procedure in order to  estimate how many of these ``past
values'' should be used: we apply an exponential fit moving from
the ``present'' towards the ``past''. For example, if the first
missing point is the $x_4$ then after the fitting it will take the
value
\begin{equation}
\label{eq:6} x_4=\frac{x_3 e^{(-\frac{1}{20})}+x_2
e^{(-\frac{2}{20})}+x_1
e^{(-\frac{3}{20})}}{e^{(-\frac{1}{20})}+e^{(-\frac{2}{20})}+e^{(-\frac{3}{20})}}
\end{equation}

Applying the exponential fit to the entire signal of
Fig.~\ref{fig:shortses1}a  we plot in Fig.~\ref{fig:shortses1}b
the signal alongside the resulting background, while in
Fig.~\ref{fig:shortses1}c we plot the final signal after
subtracting the resulting background from the original
time-series.

Next, we compute the instantaneous power \emph{P} and use it for
the analysis of the electric signal in natural time by following
the procedure described in Ref.\cite{NAT09V}. We then find
$\kappa_1=0.0626\pm0.0002$,  $S=0.0789\pm0.0002$ and
$S_-=0.0593\pm0.0003$.

An alternative way to analyse a short duration electric signal in
natural time and evaluate the parameters $\kappa_1$, $S$ and
$S_-$, is by approximating the signal under consideration with
another one which is of dichotomous nature. Along this line we do
the following: First, we rescale the signal of
Fig.~\ref{fig:shortses1}c so as its values are in the region [0,1]
according to the relation:
\begin{equation}
\label{eq:7} data=\frac{data-min(data)}{max(data)-min(data)}
\end{equation}
We then determine a threshold $V_{thres}=0.5$ for the vertical
axis designating the measured values of the potential difference
$V$ and, starting from the beginning of the signal, we compare
each value of $V$ with $V_{thres}$. All the points for which $V$
exceeds $V_{thres}$ correspond to the high-level states and the
value 1 is assigned; to all the other points we assign the value
0. Thus, we have a new, dichotomous signal consisting of
subsequent aces and zeros. The duration $Q_k$ of each pulse will
now be the sum of the consecutive aces. The dichotomous signal
along with the original signal are depicted in
Fig.~\ref{fig:dichotomous}. The analysis in natural time gives
$\kappa_1=0.062\pm0.006$, $S=0.080\pm0.007$ and
$S_-=0.057\pm0.007$. These values, after considering that the
minimum $\kappa_1$ value measured\cite{SPRINGER} to date for an SES
activity is $\kappa_1=0.063\pm0.003$, indicate that they obey
Eqs.~\eqref{eq:3} and \eqref{eq:5}, thus this signal can be
classified as an SES activity. This conclusion is consistent with
the one obtained above after using the instantaneous power $P$.

\section{Application of the Huang-Hilbert transform to an ses of short duration}\label{hhapp}

\begin{figure}
\includegraphics[scale=0.5]{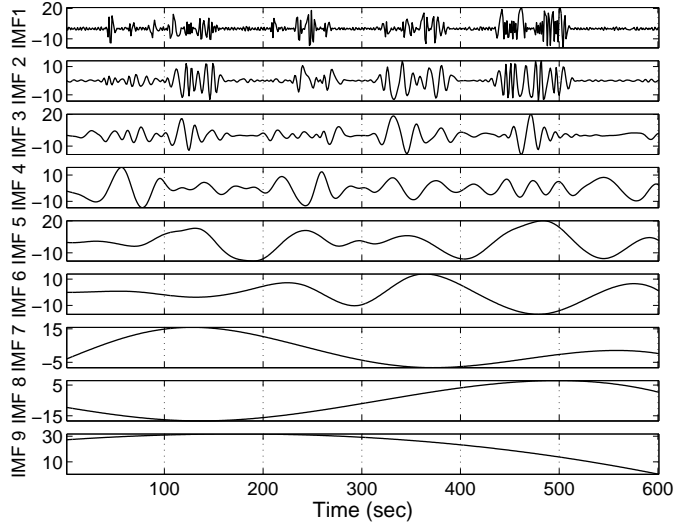}
\caption{\label{fig:imfs}The nine IMFs extracted after the sifting
process was applied to the signal in Fig.~\ref{fig:shortses1}a.
IMF9 represents the trend of the signal.}
\end{figure}

In Fig.~\ref{fig:imfs} we depict the IMFs that were extracted by
applying the procedure described in Section \ref{hh} to the signal
depicted in Fig.~\ref{fig:shortses1}a.

The first IMF is actually the oscillation with the highest
frequencies and we want to determine if it contains information
about the signal. Thus, we subtract it from the original
time-series (Fig.~\ref{fig:shortses1}a) and analyse the difference
in natural time after making this new signal dichotomous using the
method described in Section \ref{nt}. We find
$\kappa_1=0.061\pm0.006$, $S=0.079\pm0.005$ and
$S_-=0.056\pm0.009$ while during the analysis in natural time of
the same signal after eliminating the magnetotelluric background
using the exponential fit, we found as mentioned in Section
\ref{nt} $\kappa_1=0.062\pm0.006$, $S=0.080\pm0.007$ and
$S_-=0.057\pm0.007$. Thus, we conclude that by subtracting the
first IMF we can still classify the signal under consideration as
SES, thus probably indicating that the first IMF does not
contribute to the signal.

In order to examine which of the IMFs better represent the SES, we
analyse in natural time various sums of IMFs using the
corresponding dichotomous signals. The results are shown in
Table ~\ref{tab:sumsimf}. Comparing the results in
Table ~\ref{tab:sumsimf} with those of Section \ref{nt} we can see
that the better representation of the SES comes after summing IMFs
2-9, i.e. by subtracting only IMF1 from the original signal.

\begin{figure}
\includegraphics[scale=0.5]{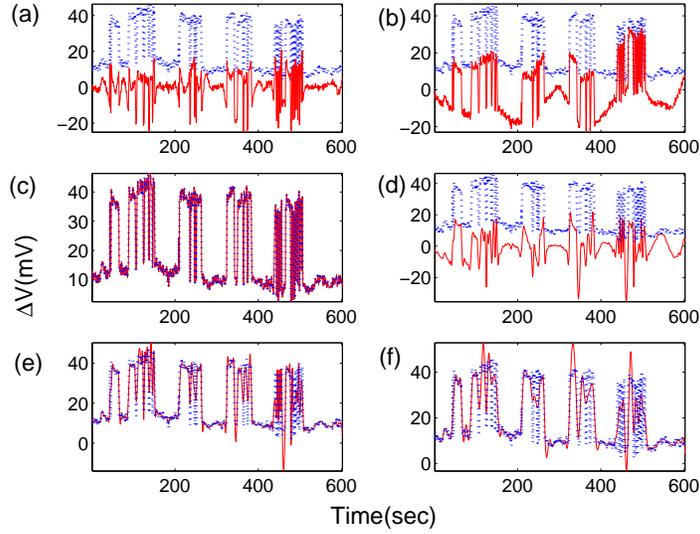}
\caption{\label{fig:sumsimf} (color online) The original
time-series (blue dotted line) and the sums of IMF
a)1-3,b)1-5,c)1-9,d)2-4,e)2-9,f)3-9 (red line)}
\end{figure}

In Fig.~\ref{fig:sumsimf} we depict the original time-series (blue
dotted line) alongside the IMF sums (red line). Their comparison
shows that complete concurrence exists only in
Fig.~\ref{fig:sumsimf}c where all the IMFs are summed up. The
original time-series  already had a magnetotelluric background so
the sum of all the IMFs, which in reality represents the original
time-series, contains a background as well. The two plots better
coincide in Fig.~\ref{fig:sumsimf}e regarding the pulses of the
signal. In this case the sifting process produces some erroneous
estimations in the values of the extrema with that near the 750sec
mark being the most characteristic. This problem might be due to
the use of the cubic spline used for the extraction of the IMFs as
described in Section \ref{hh}. The cubic spline can add large
swings near the ends of the signal which can spread inwards and
corrupt the data especially the low-frequency components. This
problem though, still allows for the sifting process to extract
the essential scales from the data\cite{huang1998} and does not affect
the basic characteristics of the signal.

\section{Analysis of the instantaneous amplitudes of the Hilbert transform}\label{instamp}

The squared instantaneous amplitude ${A(t)}^2$ of the sum of all
the IMFs can be considered proportional to the instantaneous power
$P$ of the original time-series, i.e. $P=|x(t)|^2={x(t)}^2$. From
Eq.~\eqref{eq:10}, we have $x^2={A(t)}^2-{\hat{y}}^2$. Thus, we
derive the following interrelation between $P$ and $A(t)$:
\begin{equation}
\label{eq:12}
P={A(t)}^2-{\hat{y}}^2
\end{equation}

\begin{figure}
\includegraphics[scale=0.5]{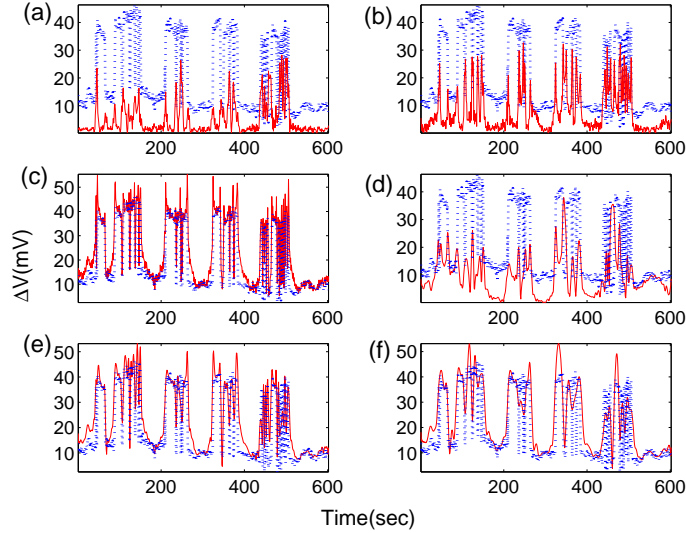}
\caption{\label{fig:sumsA} (color online) The original time-series
(blue dotted line) and the instantaneous amplitudes of the sums of
IMF a)1, b)1-3, c)1-9, d)2-4, e)2-9, f)3-9 (red line).}
\end{figure}

Our scope is to investigate whether the instantaneous amplitudes
can be used instead of the instantaneous power for the analysis in
natural time. The process is the following: We sum the IMFs and by
applying the Hilbert transform we find the instantaneous amplitude
of each sum. We then square the instantaneous amplitude and begin
the analysis in natural time by applying the method of the
instantaneous power as described in Ref.\cite{NAT09V}, treating
the squared instantaneous amplitude as the instantaneous power. We
test various IMF sums in order to examine which of the amplitudes
better represent the SES. The results are shown in
Table~\ref{tab:sumsA} while in Fig.~\ref{fig:sumsA} we depict the
original time-series alongside the instantaneous amplitudes of the
IMF sums.

An inspection of Fig.~\ref{fig:sumsA} reveals that the
instantaneous amplitude of the IMF1-9 sum, i.e., the amplitude of
the Hilbert transform of the original time series, better
approximates the original signal.

We remind that in Section \ref{nt} for the same signal using the
instantaneous power method we had found
$\kappa_1=0.0626\pm0.0002$, $S=0.0789\pm0.0002$ and
$S_-=0.0593\pm0.0003$. Comparing these results with the ones of
Table~\ref{tab:sumsA} we find that we have better results by
adding to the computations the highest-frequency component (i.e.
IMF1). Thus, the $A^2$ of the original time-series (i.e. the $A^2$
of the sum of all the IMFs) can be used for the analysis in
natural time instead of the instantaneous power.

\section{Summary and concluding remarks}\label{conclusion}

\begin{table}
\caption{\label{tab:sumsA}Analysis in natural time of the
instantaneous amplitudes of IMF sums.}
\begin{tabular}{cccc}
\hline
Sums of IMF: & $\kappa_1$ & $S$ & $S_-$ \\
\hline
1 & $0.078\pm0.005$ & $0.064\pm0.008$ & $0.064\pm0.009$\\
1-3 & $0.067\pm0.002$ & $0.068\pm0.003$ & $0.071\pm0.004$\\
1-9 & $0.062\pm0.006$ & $0.083\pm0.008$ & $0.060\pm0.006$\\
2-4 & $0.0601\pm0.0009$ & $0.0400\pm0.0008$ & $0.0418\pm0.0005$\\
2-9 & $0.067\pm0.007$ & $0.081\pm0.006$ & $0.063\pm0.007$\\
3-9 & $0.082\pm0.005$ & $0.083\pm0.005$ & $0.083\pm0.007$\\
\hline
\end{tabular}
\end{table}

Table~\ref{tab:concl} summarizes the results of the analysis in
natural time after the  application of the Huang-Hilbert transform
to an SES activity and compares them to those obtained from the
natural time analysis after eliminating the background from the
original time-series. First, we showed that IMF1 does not affect
the durations of the pulses of the signal since, by subtracting it
from the original time-series, the natural time parameters remain
the same. This might indicate that IMF1 contains information about
the distance of the source emitting the electric signal, from the
recording station. Second, we showed that the $A^2$ of the sum of
all the IMFs (i.e. the $A^2$ of the original time-series) can be
used instead of the instantaneous power of the time-series (after
having subtracted the background) for the analysis in natural time
in order to classify the signal as an SES.

\begin{table*}
\caption{\label{tab:concl}Summary of the results of the
analysis in natural time.}
\begin{tabular}{cccc}
\hline
Analysis in natural time of: & $\kappa_1$ & $S$ & $S_-$ \\
\hline
\emph{Signal after eliminating the background (dichotomous)} & $0.062\pm0.006$ & $0.080\pm0.007$ & $0.057\pm0.007$\\
\emph{Sum of IMF2-9 (dichotomous)} & $0.062\pm0.014$ & $0.078\pm0.012$ & $0.056\pm0.016$\\
\emph{Signal after eliminating the background (using $P$)} & $0.0626\pm0.0002$ & $0.0789\pm0.0002$ & $0.0593\pm0.0003$\\
\emph{A of sum of IMF1-9 (using $A^2$)} & $0.062\pm0.006$ & $0.083\pm0.008$ & $0.060\pm0.006$\\
\hline
\end{tabular}
\end{table*}


\end{document}